\documentclass{aa}

\usepackage{psfig}

\def\src  {SGR~1806--20~}

\begin{document}

   \title{Spectral evolution of weak bursts from
    SGR 1806--20  observed with INTEGRAL\thanks{Based on observations with INTEGRAL, an ESA project with instruments and science data centre funded by ESA member states (especially the PI countries: Denmark, France, Germany, Italy, Switzerland, Spain), Czech Republic and Poland, and with the participation of Russia and the USA.} }

   \author{D. G\"{o}tz\inst{1,2}, S. Mereghetti\inst{1},
    I.F. Mirabel\inst{3,4}, \and K. Hurley\inst{5}
    }

   \offprints{D. G\"{o}tz, email: diego@mi.iasf.cnr.it}

   \institute{Istituto di Astrofisica Spaziale e Fisica Cosmica -- CNR,
              Sezione di Milano ``G.Occhialini'',
          Via Bassini 15, I-20133 Milano, Italy
         \and
             Dipartimento di Fisica, Universit\`{a} degli Studi di Milano Bicocca,
             P.zza della Scienza 3, I-20126 Milano, Italy
         \and
         Service d'Astrophysique, CEA/Saclay, Orme des Merisiers B\^{a}t. 709, F-91191 Gif-sur-Yvette, France
         \and
         Instituto de Astronomia y Fisica del Espacio / CONICET, cc67, suc 28. 1428 Buenos Aires, Argentina
     \and
         UC Berkeley Space Sciences Laboratory, Berkeley CA 94720-7450, USA
         }


\abstract{
We report on  bursts from the Soft Gamma-Ray Repeater
SGR 1806--20 detected with   INTEGRAL  in
October 2003, during a period of moderate activity of the source.
The spectral and temporal properties of 21 short bursts are consistent
with those found in previous observations, even if these bursts are among the faintest
observed in the 15-200 keV range from this source.
During some of the bursts a clear spectral evolution is visible.
The data also show, for the first time, evidence for a  hardness-intensity
anti-correlation within SGR 1806--20 bursts.
}

\authorrunning{D. G\"{o}tz et al.}

\maketitle

\section{Introduction}

Soft Gamma-ray Repeaters (SGRs) are a class of peculiar high-energy sources discovered 
through their recurrent emission of soft $\gamma$-ray bursts.
These bursts have typical durations of $\sim$0.1 s 
and luminosities in the range 10$^{39}$-10$^{42}$ ergs s$^{-1}$
(see \cite{hurleyrew} for a review of this class of objects). The bursting
activity and the persistent emission observed in the $\sim$0.5-10 keV energy
range are generally explained
in the framework of the ``Magnetar'' model (e.g. \cite{dt92}, \cite{td95}), as caused by
a highly magnetized ($B\sim$10$^{15}$ G) slowly rotating ($P\sim$ 5-8 s) neutron star.

SGR 1806--20 is one of the most active Soft Gamma-ray Repeaters.
Here we report new observations of this source obtained with the INTEGRAL satellite
in October 2003 during a period of bursting activity
(\cite{gotza}, \cite{hurley}, \cite{merec}, \cite{gotzb}).
These data have two advantages compared to previous observations
in the soft $\gamma$-ray energy range of bursts from this source.
First, they have been obtained with an imaging instrument, thus we can
exclude that the bursts originate from a different source in the field.
Second, they have a good sensitivity and time resolution which allows us to study
the spectral evolution of relatively faint  bursts.

\begin{figure*}
      \hspace{0cm}\psfig{figure=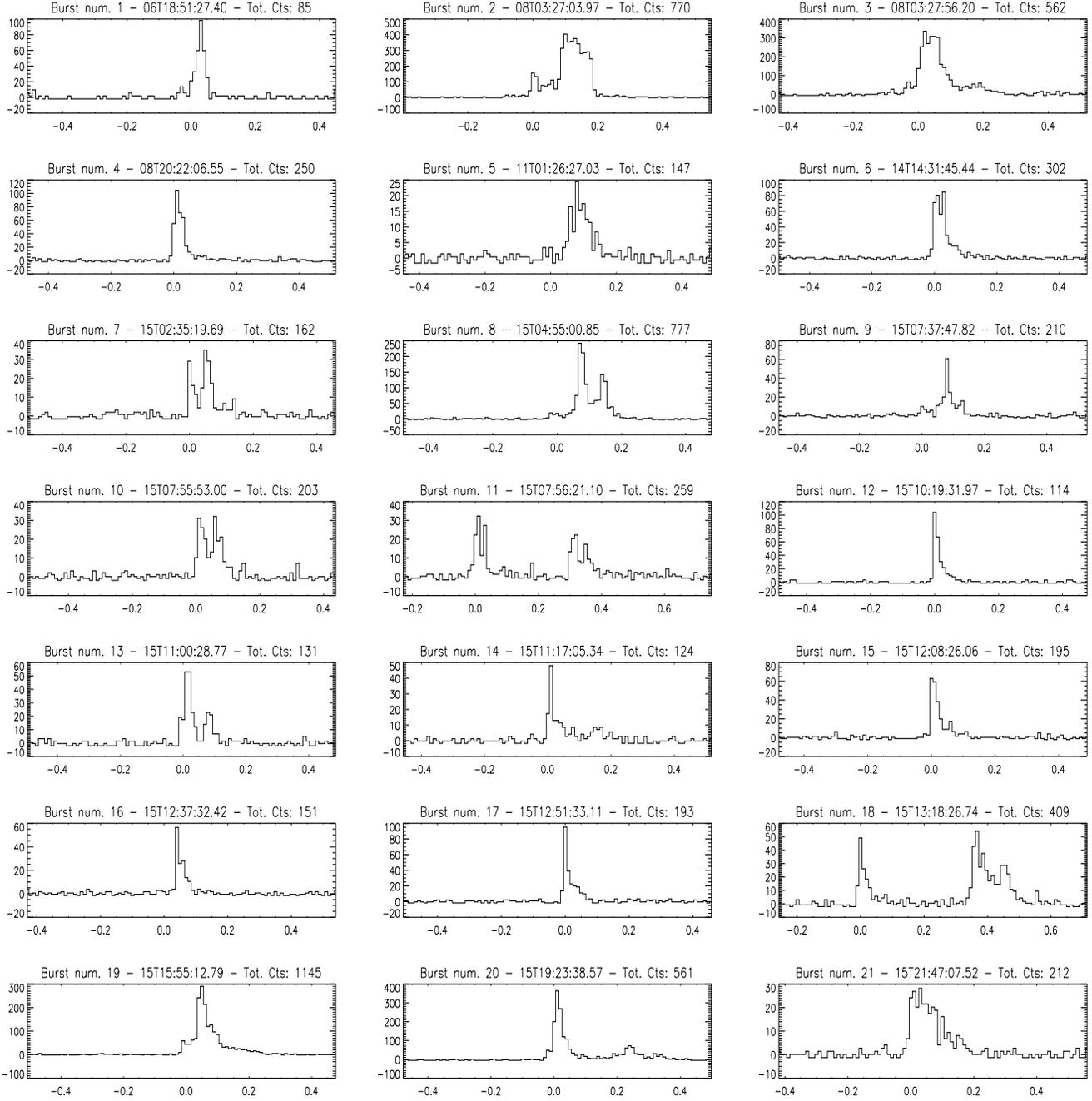,width=18cm,height=18cm,angle=90}
      \caption[]{IBIS/ISGRI background subtracted light curves of
the SGR 1806-20 bursts in the 15-100 keV range.
Each panel corresponds to a time interval of one second
and the time bins are of 10 ms.
Units of the axes are time in seconds and vignetting corrected counts per bin.
Time 0 corresponds to the
the starting time of the $T_{90}$ computation and is reported on top of each panel
together with the total number of net counts.}
         \label{lc}
   \end{figure*}

\section{Observations and data analysis}

The region of \src was observed by INTEGRAL (\cite{winkler}) between October 8 and 15 2003
as part of the Core Program deep observation of the Galactic Centre,
yielding an exposure of about $\sim$480 ks on the source.
Several bursts from the direction of \src were detected in near real
time by the INTEGRAL Burst Alert System (IBAS, \cite{ibas}), using data
from the IBIS instrument (\cite{ibis}).
IBIS, a coded mask telescope with a large field of view
(29$^{\circ}\times29^{\circ}$),  comprises
two detector layers: ISGRI (15 keV - 1 MeV, \cite{isgri})
and PICsIT (170 keV - 10 MeV, \cite{picsit}).
Only ISGRI data are relevant here, since PICsIT does not have enough time resolution
for the study such short bursts.

In total, 21 bursts were detected by the IBAS programs.
By means of images accumulated over the time intervals corresponding to the
individual bursts, we can be confident that all of them originated from \src.
In fact, the derived coordinates are all within 2$'$ from
the well known position of SGR 1806-20 (\cite{chandra}), while the 90\%
confidence error circle is typically of the order of 2.5$'$.
In particular, the bursts positions are not consistent with the possible
SGR 1808-20 (\cite{lamb}) recently discovered at 15$'$ from \src.

\begin{figure*}
     \hspace{0cm}\psfig{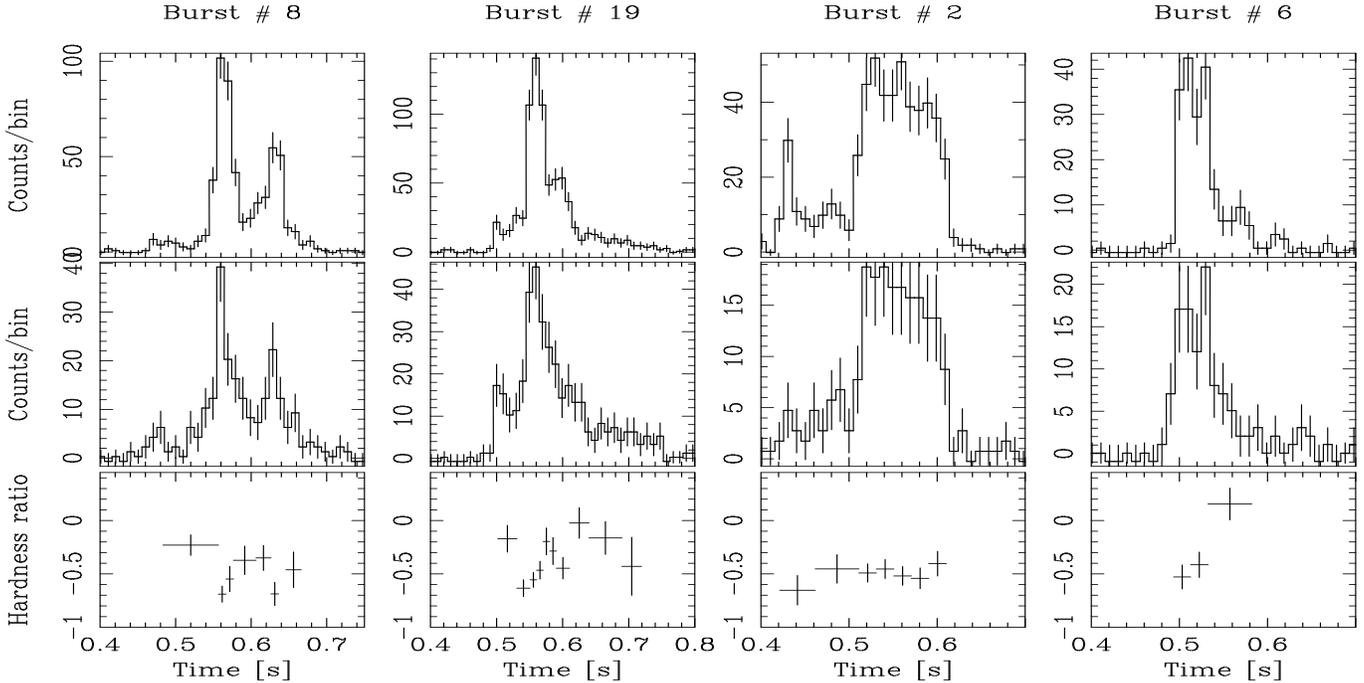}
      \caption[]{15-40 keV light curve (\emph{Top Panels}), 40-100 keV light curve
      (\emph{Middle Panels}), time resolved hardness ratio (\emph{Bottom Panels})
      for four bursts with good statistics. The time resolved hardness ratio for
      bursts number 8,19,6 is inconsistent with a constant value at $\sim$3.5 $\sigma$ level.
      }
         \label{trehr}
\end{figure*}

The background subtracted light curves of the bursts, binned at 10 ms,
are shown in Fig. \ref{lc}.
In order to increase the signal-to-noise-ratio, they were extracted from ISGRI
pixels illuminated by the source for at least half of their surface
and selecting counts in the  15-100 keV energy range (most of the bursts
had little or no signal  at higher energy).
The bursts were detected at various off-axis angles, ranging
from 2.5 to 13.3 degrees, corresponding to a variation of 80\% in the instrument
effective area.
The light curves shown in Fig. \ref{lc} have been corrected for this
vignetting effect.  The total number of net counts actually recorded for each burst
is indicated in the corresponding panel.

The light curves shown in Fig. \ref{lc} have  shapes  typical for SGR bursts.
From the light curves we determined the $T_{90}$ duration of each burst (i.e. the
time during which 90\% of the total burst counts are accumulated).
The $T_{90}$ values
range typically from $\sim$0.1 to $\sim$0.2 s for single
peaked bursts and can be as long as $\sim$0.6 seconds for double
peaked bursts. In fact the $T_{90}$ values of these bursts include
the ``interpulse'' period.
Some bursts are preceded by a small  precursor.

The peak flux  and fluence for each burst were first derived in counts units
from the light curves of Fig.\ref{lc}, and then converted to physical units
adopting a constant conversion factor derived from the spectral analysis of the
brightest bursts (see next section).
The resulting 15-100 keV peak fluxes and fluences are respectively  in the range
(4--50)$\times10^{-7}$ erg cm$^{-2}$ s$^{-1}$ ($\Delta t$=10 ms)
and (2--60)$\times10^{-8}$ erg cm$^{-2}$.
Within the large uncertainties, the fluence distribution is consistent
with the power law slope found by G{\" o}{\u g}{\" u}{\c s} et al. (2000). 
Many of these bursts are among the faintest
ever detected from  SGRs at these energies.

\subsection{Spectral properties}

For the  bursts with more than 500 net counts we could perform a detailed
spectral analysis.
The 15-200 keV spectra, integrated over the whole duration of each burst,
were well fitted by an Optically
Thin Thermal Bremsstrahlung (OTTB) model, yielding
temperatures in the range from 32 to 42 keV.
We tried other  models, like a
power law or a black body, but they
were clearly ruled out.

Adopting a temperature $kT$=38 keV (consistent with the average
spectra of the brightest bursts) we derived a conversion factor of
1 count s$^{-1}$ = 1.5$\times$10$^{-10}$ erg cm$^{-2}$ s$^{-1}$
(15-100 keV), which we adopted for all the bursts.

To investigate the time evolution of the burst spectra we computed
hardness ratios, defined as  $HR=(H-S)/(H+S)$, based on the
background subtracted counts in the ranges 40-100 keV ($H$) and 15-40 keV ($S$).
The time resolved $HR$ values were computed for all the bursts with more than 200 net counts
(i.e. for 12 bursts of our sample).
The duration of the individual time bins were chosen
in order to have at least 80 net counts in the total ($H+S$) band.

Some bursts show a significant spectral evolution, while
others, particularly those with a  ``flat topped'' profile, do not.
Some examples are given in  Fig. \ref{trehr}. While several bursts
show a soft-to-hard evolution (e.g. number 6), others show
a more complex evolution (eg. number 19).

\begin{figure}
     \hspace{0cm}\psfig{figure=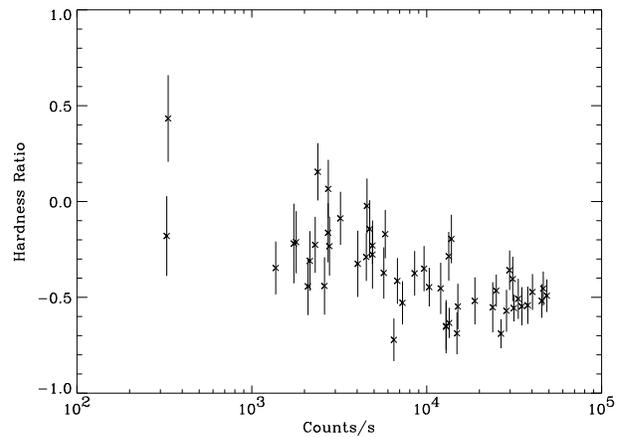,angle=90,width=8.5cm}
      \caption[]{Hardness-Intensity plot of the time resolved hardness ratios of
the 12 bursts with the best statistics.
The hardness ratio is defined as $(H-S)/(H+S)$, where $H$ and $S$ are the
background subtracted counts in the ranges 40-100 keV and 15-40 keV respectively.
The count rates are  corrected for the vignetting and refer to the 15-100 keV range.
}
         \label{hi}
   \end{figure}

We investigated the variation of the hardness ratio
versus intensity ($I$). Considering all the time bins of all the bursts (see Fig. \ref{hi}),
we find a hardness-intensity anti-correlation.
The linear correlation coefficient of the data plotted in Fig. \ref{hi}
has a chance probability $P$ smaller than 10$^{-3}$ of being due
to uncorrelated data. According to an F-test,
the data are significantly ($\sim$5.2 $\sigma$) better described
by a linear fit ($HR = 0.45 - 0.22\times log(I)$) than by a constant value.
The exclusion of the three ``flat topped'' bursts
from the fit does not affect the statistical significance of the
anti-correlation.

We also find an hardness-fluence anti-correlation over the entire fluence
range of our bursts, although with a smaller statistical significance 
($P$ = 5$\times$10$^{-3}$), which is consistent with our hardness-intensity
anti-correlation and also with the results obtained with {\it RXTE} 
data at lower energies (\cite{gogus}).

\section{Discussion}
\label{disc}

Previous studies indicated weak  or no
spectral evolution for SGR bursts
(e.g. \cite{fenimore}, \cite{kouveliotou}).
Up to now indication for a  hard-to-soft evolution has
been reported only for a single burst from \src  (\cite{strohmayer}) and
for a precursor to a long  burst (3.5 s) from SGR 1900+14 (\cite{ibrahim}).
The same kind of spectral evolution has also been 
reported for a $\sim$9 s long burst from SGR 0526-66 (\cite{golenetskii}):
the softening trend is seen in the first three of the four spectra extracted,
while the last one is as hard as the first one.
In our sample we do not find evidence for this kind of evolution.

On the other hand, soft-to-hard evolution has been seen with the BATSE instrument
for two peculiar bursts very likely originating from
SGR 1900+14 (\cite{woods}).
These two bursts were quite different from the usual bursts, both in terms of duration
(lasting $\sim$ 1 s), and spectral  hardness ($kT$ of the
order of 100 keV).

In the framework of the magnetar model (\cite{dt92}),
short ($\sim$0.1 s) SGR bursts are
usually  described as the radiation originating from
the cooling of an optically thick pair-photon plasma. This plasma is
generated in the neutron star  magnetosphere by an Alfv\'{e}n
pulse, which is triggered by a sudden shift in the magnetospheric
footpoints driven by a fracture in the neutron star crust (\cite{td95}).
This model is able to explain the time histories and energetics of the typical
SGR bursts, and predicts that the effective temperature of the spectra
should vary weakly during the bursts,
owing to the constancy of the photospheric energy flux.
No detailed predictions are available concerning more complex spectral evolution
as we observe in some of the bursts. For example
the model does not account for the presence of a soft precursor as observed
in burst number 2 (see Fig. \ref{trehr}).

Our results  indicate that a hardness-intensity
anti-correlation (which in many bursts manifests itself as a soft-to-hard
time evolution) is present in  bursts from \src which are not particularly long, nor
spectrally hard and not at all very energetic. It is interesting to note that
this correlation is opposite to what found for the bursts emitted from
1E 2259+586 (\cite{gavrill}), which although have lower fluences than the ones we measure.

\begin{acknowledgements}
This research has been supported by the Italian Space Agency.

\end{acknowledgements}

\end{document}